\documentclass{emulateapj}
\usepackage{times}
\usepackage{lscape}

\usepackage[pdftex]{color}

\def\la{\mathrel{\mathpalette\fun <}}
\def\ga{\mathrel{\mathpalette\fun >}}
\def\simpropto{\lower.2ex\hbox{$\; \buildrel \sim    \over \propto \;$}}

\def\simpropto{\lower.2ex\hbox{$\; \buildrel \sim \over \propto \;$}}

\def\fun#1#2{\lower0.837ex\vbox{\baselineskip0ex\lineskip0.209ex
  \ialign{$\mathsurround=0ex#1\hfil##\hfil$\crcr#2\crcr\sim\crcr}}}

\def\msun{M_\odot}
\def\msunyr{M_\odot \ {\rm yr}^{-1}}
\def\msunsec{M_\odot \ {\rm s}^{-1}}

\def\sles{\lower2pt\hbox{$\buildrel {\scriptstyle <}
   \over {\scriptstyle\sim}$}}

\def\sgreat{\lower2pt\hbox{$\buildrel {\scriptstyle >}
   \over {\scriptstyle\sim}$}}

\def\la{\mathrel{\mathpalette\fun <}}
\def\ga{\mathrel{\mathpalette\fun >}}

\def\msun{M_\odot}
\def\msunyr{M_\odot \ {\rm yr}^{-1}}
\def\msunsec{M_\odot \ {\rm s}^{-1}}

\begin{document}

\title{Fall-Back Disks  in Long and Short GRBs}
\shortauthors{CANNIZZO, TROJA, \& GEHRELS}
\author{
 J.~K.~Cannizzo\altaffilmark{1,2},
       E.~Troja\altaffilmark{3,4}, 
  and
      N.~Gehrels\altaffilmark{4}}

\altaffiltext{1}{CRESST and Astroparticle Physics Laboratory 
                   NASA/GSFC, Greenbelt, 
                    MD 20771;
               John.K.Cannizzo@nasa.gov}
\altaffiltext{2}{Department of Physics, 
                 University of Maryland, Baltimore County, 
                 1000 Hilltop Circle, Baltimore, MD 21250}
\altaffiltext{3}{NASA Postdoctoral Program Fellow}
\altaffiltext{4}{Astroparticle Physics Laboratory NASA/GSFC, 
                 Greenbelt, MD 20771}

\begin{abstract}
We  present  numerical time-dependent  calculations
   for fall-back disks  relevant to GRBs in which the disk of material
surrounding the black hole (BH) powering the GRB jet
modulates the mass flow, and hence the strength of the jet.
 Given the initial existence of a small mass $\la 10^{-4}\msun$
near the progenitor  with a circularization radius
   $\sim10^{10} - 10^{11}$ cm,  an unavoidable consequence will be
  the formation of an ``external disk''  whose outer edge
   continually moves to larger radii due to  angular momentum transport
    and lack of a confining torque.
     For long GRBs, if the mass distribution  in the initial
   fall-back disk traces the progenitor envelope, then a
  radius $\sim 10^{11}$ cm gives a time scale $\sim 10^4$ s
for the X-ray plateau. For late times $t>10^7$ s a steepening
  due to a cooling front in the disk may have observational 
  support in GRB 060729.
   For short GRBs, one expects  most of the mass initially to lie at 
  small radii $<10^8$ cm;
   however  the presence of even a trace amount $\sim 10^{-9}\msun$
   of high angular material can give a brief plateau in the light curve.
    By 
   studying the plateaus in the
  X-ray decay of GRBs, which can last 
up to $\sim10^4$ s after the prompt
  emission, 
  Dainotti et al. find
   an apparent inverse relation between the 
  X-ray luminosity at the end of the plateau and 
the duration of the plateau.
   We show that this relation may simply represent the 
fact that one is biased against
detecting faint plateaus, 
 and therefore preferentially sampling the more energetic
  GRBs. If, however, there were a standard 
  reservoir in fall-back mass, our model 
can reproduce the inverse X-ray luminosity-duration relation.
      We emphasize that we do not
address the very steep, initial decays immediately following
  the prompt emission, which have been modeled by Lindner et al.
 as fall-back of the progenitor core, and may entail the accretion
 of $\ga 1\msun$.
\end{abstract}


\section{Introduction}

 The tentative pre-{\it Swift}
    hints that breaks
   occur in the long-term afterglow light curves
   simultaneously
   in different wavebands at $\sim1-10$ d (Frail et al. 
   2001)
   after the GRB
have not been borne out by numerous detailed observations
of {\it Swift}  GRB afterglows in recent years
(Ghisellini et al. 2007,
      Oates et al.  2007, 
    Racusin et al. 2008,
  Liang et al. 2008).
  With {\it Swift} (Gehrels et al. 
    2004), the long-term
($t\ga 10^6$ s) X-ray behavior has turned out to be
surprisingly complex, exhibiting alternating steep and shallow
  slopes when plotted as log $F_x$ versus log$(t-T0)$,
 where $T0$ is the trigger time for prompt BAT emission
   (Zhang et al. 2006;
   Nousek et al. 2006).


The traditional viewpoint that long-term fading
  of the GRB afterglow represents the deceleration
 and spreading of a relativistic jet (e.g., Sari, Piran, Halpern
   1999, Frail et al. 2001)
 has given way to a much broader effort involving
  a diverse variety of hypotheses by which a central
  engine can be powered for a long time,
  including such scenarios as magnetars 
  (Fan \& Xu 2006;
    Toma et al. 2007;
   Rowlinson et al. 2010),
   quark stars (Staff, Niebergal, \& Ouyed 2008),
 and long term accretion onto a BH formed during the
collapsar or NS-NS merger
     (Kumar et al. 2008a, 2008b;
    Metzger et al. 2008a, 2008b).
   The accretion hypothesis
   entails
  exploring the possibility that the
long-term decay of the X-ray flux is not due to the deceleration
of  baryonic ejecta, but rather a secular decrease in the
rate of accretion powering the central engine,
  and therefore indirectly   the jet
   (Kumar et al. 2008a,
                 2008b;
        Metzger et al. 2008a, 2008b, 
 Cannizzo \& Gehrels 2009, 
           hereafter  CG09). 
 For long GRBs (lGRBs),
    the early, steep rate of decay may be giving us information
about the radial density distribution within the progenitor
core (Kumar et al. 2008b), 
  whereas the later decay may be governed by the outward
expansion of the transient disk formed from the remnants
  of the progenitor
              (CG09).
    As regards short GRBs (sGRBs),
   even if a small
  amount of material ($\sim 10^{-5} - 10^{-4}\msun$) is expelled during
the NS-NS merger and later accreted in a disk,
that would be sufficient to power a bright afterglow,
  which may also be strongly influenced by the effects of r-process
nucleosynthetic heating in the neutron rich material that becomes 
the disk (Metzger et al. 2010).


    Zhang et al. (2006) 
    present a        schematic for
 the decaying GRB light curve as seen by the XRT on {\it Swift}.
 The decay is traditionally shown in $\log F - \log t$.
 There are four basic power-law decay ($F \propto t^{-\alpha}$)
 regimes: (i) a steep decline following the prompt emission with $\alpha_{\rm I} \simeq  3$   out to $10^2-10^3$ s,
         (ii) a plateau with $\alpha_{\rm II} \simeq  0.5$ out to $10^3-10^4$ s,
        (iii) a steepening with $\alpha_{\rm III} \simeq  1.2$ out to $10^4-10^5$ s,  and
         (iv) a further steepening at late times (not always seen) with $\alpha_{\rm IV} \simeq 2$.

     CG09
    present 
    a general  analytical formalism
   to explain the different power law decays
   using 
   a fall-back disk, where the variations in $\alpha$
   could potentially be explained by different physics 
   operating within the disk. The results of 
        CG09 
   were purely analytical;  in this work we present 
   time dependent numerical calculations  in order to 
   examine in more detail the potential of the model, and we apply
   the results to XRT data for one lGRB and one sGRB,
   taking the best studied of each class.
   In Section 2 we review the Dainotti relation,
   an empirical relation involving the  duration and luminosity
   of segment II, in 
   Section 3 we present our detailed numerical model, 
   in  Section 4 we compare the model with observations for the lGRB
   060729, the GRB with the longest observational time series in X-rays,
   in Section 5  we compare theory and data for the sGRB 051221A,
   the sGRB with the longest and most detailed XRT light curve,
   in Section 6 we revisit the Dainotti relation, in the context of our
   numerical results, and
   in Section 7 we discuss and summarize our results.

\section{ The  Dainotti Relation}

 Dainotti et al. (2008, 
                  2010) found an empirical relation
 between the duration of the X-ray plateau 
   in the source frame
     $t^*_{\rm II}=
        t_{\rm II} (1+z)^{-1}$,
and the X-ray luminosity
        at the end of the plateau,
  also corrected into the source frame,  $L^*_{\rm II}$.
Expressing their relation in the form
 $\log_{10} L^*_{\rm II} ({\rm erg} \ {\rm s}^{-1})
   =  a + b \log_{10} t^*_{\rm II}({\rm s})$,
Dainotti et al. (2010) find $a=51.1\pm1$
and $b=-1.1\pm0.3$.
  Their sample of lGRBs with small errors
  is defined by $u<4$, 
   where $u\equiv
      [(\log \delta {L^*_{\rm II}})^2 
     + (\log \delta {t^*_{\rm II}})^2]^{1/2}
      $, 
   and $\log \delta {L^*_{\rm II}}$
   and $\log \delta {t^*_{\rm II}}$
represent the logarithmic errors in $L^*_{\rm II}$
 and $t^*_{\rm II}$, respectively. 
  Using
 the  same sample of 62 lGRBs 
     with small errors
 taken from  Dainotti et al. (2010),
we find\footnote{
 We employ  a different method 
   than  Dainotti et al. 
   (2010),
  a Monte Carlo technique in which 
 $10^6$ data sets are created with  $1\sigma$ errors
   randomly either added or subtracted to each data point
  in
   both $x$ and $y$ directions. Thus
       for $N$ data points
   one could in principle have $N^4$ distinct data sets.
  For each data set,
    $a$ and $b$ values are calculated,
  and the final values for $a$, $\delta a$, $b$, and $\delta b$
   are taken from the averages and standard deviations
   of the $10^6$ values.}
     $a=50.5\pm0.4$ and  $b=-0.92\pm0.1$,
   close to their results.

\vskip -0.065 cm
\begin{figure}[h!]
\begin{center}
\includegraphics[height=130mm]{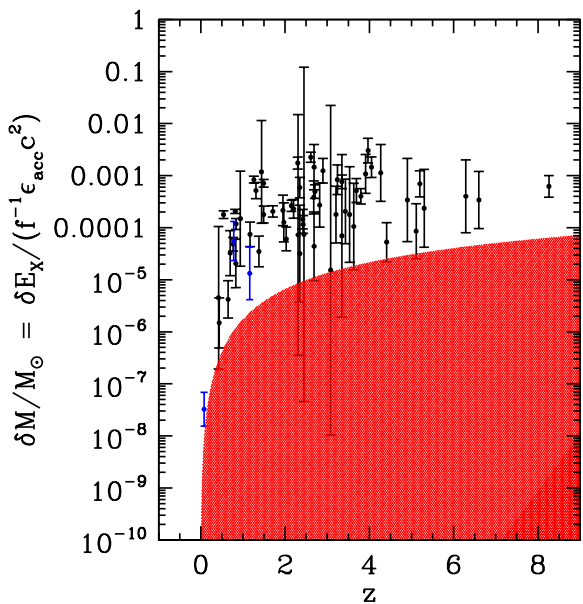}
\end{center}
\vskip -4.125 cm
\caption{
   Inferred total accretion mass for the plateau $+$ later decay phases
   of GRBs, using data from Dainotti et al. 
       (2010) for their 62 
   well-constrained lGRBs.
   The hatched area (shown in red)
  indicates a  putative limiting XRT detection 
 flux  level $f_{\rm II} = 10^{-12}$ erg cm$^{-1}$ s$^{-1}$
  (adopting a plateau duration $t_{\rm II} = 10^4$ s)
   for being able to study a plateau  
        to sufficient accuracy that it would have been included
   in the Dainotti et al. sample of 62 GRBs with good statistics
  and known 
    redshifts. (Three of the redshifts given
   by Dainotti et al.,
   $z=0.08$ for 051109B, 
   $z=0.78$ for 060202, 
  and  $z=1.16$ for 07051B,
   are less secure than the others, 
   i.e., based either on photometry
   or X-ray spectroscopy,
   and have been indicated by
   blue symbols.)
 We adopt a beaming factor $f=1/300$ 
  and a net  efficiency for powering the X-ray  flux
   $\epsilon_{\rm net}=
   \epsilon_{\rm acc}\epsilon_X=0.03$ 
 to convert from X-ray fluence
   (i.e., total X-ray energy, after taking into account
    $4\pi d_L^2$) to accreted mass.
}
\label{fig:1}
\end{figure}

If the inverse relation between 
  $L^*_{\rm II}$ and  $t^*_{\rm II}$ is physical,
 and if long-term accretion is the correct explanation for the long-term
X-ray light curve of GRBs, that would have implications
   for the amount of  mass  in the initial
  fall-back disk. To first order, the fact that $L^*_{\rm II} t^*_{\rm II}$
is constant would imply the accreted mass reservoir is constant.
  Willingale et al. (2007)
   present a simple formalism for integrating the X-ray flux on the 
plateau and subsequent $\alpha_{\rm III}\simeq 1.3$ decay to obtain 
 a total energy for segments II and III.
   We can convert this into a mass by making a few plausible assumptions
   about the energetic efficiencies
  (Krolik, Hawley, \& Hirose 2007),
  e.g.,
       an X-ray afterglow
    beaming
   factor $f\simeq 10^{-3} - 10^{-2}$,
    and an efficiency 
   $\epsilon_{\rm net} \simeq 0.01-0.1$
     with which accretion 
   onto the inner engine powers the observed X-rays, presumed to be 
   created within the beamed jet.
 This overall accretion efficiency has two
components, (1) the efficiency $\epsilon_{\rm acc}$
         with which accretion onto
  the BH powers the jet, where $\epsilon_{\rm acc}\simeq0.4$
  for the anticipated high spin BHs in GRBs,
   and (2) the efficiency $\epsilon_X$ with which 
     the jet power is converted into
$0.3-10$ keV X-rays that can be observed by XRT.
  Therefore  $\epsilon_{\rm net} = \epsilon_{\rm acc} \epsilon_X$.
  We use eqns. [3] and [4]
  from Willingale et al. (2007) to  obtain 
 a total energy $\delta E_X$ 
      and hence $\delta M$
    derived from the X-ray fluence,
 for segments II and III.
  The results are shown in Figure 1.
   The  hatched area 
  is bounded from above by
    an estimate
  of the effective   
     limiting XRT
flux\footnote{Our adopted XRT plateau detection
  level  $10^{-12}$ erg cm$^{-1}$ s$^{-1}$ 
   is unrelated to the nominal single pointing
 detection limit $\sim10^{-14}$ erg cm$^{-1}$ s$^{-1}$,
  and in fact exceeds it considerably:
   In order 
  to distinguish the different phases of the
canonical light curve (i.e., plateau versus steep decay),
  one must take into account the effect of 
 {\it Swift} orbital gaps, which places a
  much stronger constraint on the light curve characterization
than   simple flux detection.
  In addition,  the presence of flares may further complicate the 
 picture, but they usually are confined  to $t\la10^3$ s
  and   generally do  not represent a major portion
 of the  total energy budget.}  for being
  able to observe and characterize plateaus sufficiently
   well
    that they would satisfy $u<4$.
   There appears to be a selection  effect 
 giving rise to the Dainotti relation,
  namely that the flux detection limit for XRT prevents 
   one from clearly observing and parameterizing faint plateaus 
  at high $z$. The  Dainotti relation, 
     effectively equivalent to  $L^*_{\rm II} t^*_{\rm II} \simeq $ constant,
    results from the sampling of GRBs primarily beyond $z\simeq 1.5$,
            for which faint plateaus are observationally biased against.
   Thus we are seeing basically a narrow strip corresponding
 to the upper end of  a much broader distribution,
    which makes it appear  that 
  $L^*_{\rm II} t^*_{\rm II} \simeq $ constant.
   For $z\la1.5$ one sees curvature in the lower envelope of $\delta E_X$ values
   due to the strong dependence of the detection limit on $z$.
   Nevertheless, the fact that there appears to be a well-defined upper limit
  $\delta M \approx 10^{-4} - 10^{-3} \msun $
   is interesting: if one starts with a $\sim10\msun$
   progenitor and if accretion
   governs the long-term X-ray light
   curve, this indicates that no more than  
      a fraction  $\sim 10^{-5} - 10^{-4}$ of the progenitor mass
   survives the hypernova to form a fall-back disk (excluding segment I).
 Given the potential for strong outflows
  during the accretion process, this  $\delta M $
  should be viewed as a lower  limit.

\section{Accretion Disk Physics}

By writing the equations for mass continuity
and angular momentum transport
   in cylindrical coordinates,
  assuming
 Keplerian rotation  $\Omega_K^2=GM_{\rm BH} r^{-3}$
    and
integrating  over the vertical thickness of the accretion disk,
one arrives at an equation for the evolution of the surface density
$\Sigma=2\rho h$, where $\rho$ is the density and $h$
the disk semithickness (actually pressure scale height),
\begin{equation}
{\partial  \Sigma  \over \partial t} =
{3\over r} {\partial  \over \partial r}
\left[ r^{1/2} {\partial  \over \partial r} \left( \nu \Sigma r^{1/2} \right)\right].
\label{evolution}
\end{equation}
 The kinematic viscosity coefficient
\begin{equation}
\nu = {2 \alpha_{\rm SS} P  \over 3 \Omega_K \rho},
\end{equation}
where $P$ is the pressure and $\alpha_{\rm SS}$ is the Shakura-Sunyaev parametrization
of the angular momentum transport and heating (Shakura \& Sunyaev 1973).
 Guided by the current concordance between inferred values of
$\alpha_{\rm SS}\simeq0.1$ from dwarf nova outburst decays,
  both fast, thermal decays (Smak 1984) and
       slow, viscous decays (Cannizzo et al. 2010),
 and global 3D general relativistic
   magnetohydrodynamic (GRMHD) models 
   of the magnetorotational instability (MRI) 
   which also give $\alpha_{\rm SS}\simeq0.1$ 
   at large radii in the disk
              (McKinney \& Narayan 2007ab),
   we set $\alpha_{\rm SS} = 0.1$.
 Equation (1) is discretized following the method of 
  Bath \& Pringle (1981), 
   in which grid points are distributed as $r^{1/2}$. 
In addition,  one has a thermal energy equation
   governing the temperature evolution,
\begin{equation}
  {\partial T\over \partial t} =
    {2(A-B+C+D)\over {c_p \Sigma}}
  -{{\mathcal R} T\over {\mu c_p}} {1\over r} {\partial \over \partial r} (r v_r)
  -v_r {\partial T \over \partial r},
\end{equation}
   where the viscous heating $A=(9/8)\nu\Omega^2\Sigma$,
  the radiative cooling $B=\sigma T_e^4$, 
  and $C$ and $D$ represent radial heat fluxes
  due to turbulent and radiative transport (Cannizzo 
   et al. 2010).
   For the calculations we present in this work, the
   thermal equation only becomes of potential importance
 at very late times $>1$ yr.
 In other words, $A=B$ to very good precision
  for most of the evolution, due to the fact that
$\zeta \equiv d\log T(\Sigma)/d\log \Sigma > 0$. 
    In other words, disks with $\zeta > 0$ are viscously
 and thermally stable (Piran 1978), 
   and therefore simple viscous evolution
provides a satisfactory   physical description.

  After $\sim1$ yr the surface density in the very outer disk
  drops below that associated with the hysteresis due to
the transition between ionized and neutral gas,
the source of the dwarf nova limit cycle mechanism.
This launches a cooling front that propagates to smaller radii
and rapidly shuts off the supply of gas to the inner disk, 
as the effective viscosity  plummets.
  The dwarf nova limit cycle
  is an accretion disk instability 
  in which the $T(\Sigma)$ relation
   exhibits a hysteresis at a temperature
 corresponding roughly to the peak in the Rosseland
  opacity curve $-$ effectively an
``S''-curve when plotted as $T$ versus $\Sigma$.
    It accounts for the outbursts
  seen in dwarf novae (Lasota 2001).
      The global manifestation of
  the instability is that
   material accumulates in the disk during quiescence
  and accretes onto the central star during outburst.
  The transition between the two states is mediated
  by the action of heating and cooling fronts.
The cooling front we see at late times 
   in our calculations are a result of the disk mass
 becoming so small that the transition from 
ionized to neutral gas is instigated.
  (A detailed discussion of the effects associated 
with the heating and cooling fronts is given 
  in Cannizzo [1993, 1998].)

   At early times and small radii
   the local disk accretion can be
   significantly super-Eddington,  and in the
thin disk formalism one would have $h/r >> 1$,
 an inconsistency. 
   Section 2.3.2 of CG09 contains scalings
for the slim disk branch of solutions (Abramowicz et al. 1988)
  which we employ when $h/r > (3/4)^{1/2}$ locally.
  Combining eqns. (24) and (26) of CG09 gives
\begin{equation}
 T(\Sigma)  = 1.56 \times 10^5 \ {\rm K} \ 
        m_{\rm BH,1}^{1/4}
        r_{11}^{-1/2}
       \Sigma^{1/4},
\end{equation}
 where $m_{\rm BH,1}=M_{\rm BH}/(10\msun)$ and
    $r_{11} = r/(10^{11} \ {\rm cm})$.
  For completeness, we note that at early times and small radii,
  other types of disks are possible, such as neutrino-cooled
  accretion disks 
   (e.g., Narayan, Piran, \& Kumar 2001;
         Kohri, Narayan, \& Piran 2005;
            Chen \& Beloborodov 2007).
 The controlling dynamics of the long term accretion rate is
  governed by the physical state of the outer disk, since that
is where the slowest (i.e., controlling) time scale is.
    A complete physical modeling of the disk at small radii,
 including neutrino cooling and nucleosynthesis,
 is beyond the scope of this work, but also not necessary
  given in our  primary interest, the long term accretion rate. 
  At late times and large radii where $h/r < (3/4)^{1/2}$,
   we utilize the thin disk scalings from CG09.
    Combining eqns. (10) and (11) from CG09 gives
\begin{equation}
 T(\Sigma)  = 84.1  \ {\rm K} \ 
        \alpha_{\rm SS,-1}^{1/3}
         m_{\rm BH,1}^{1/6}
        r_{11}^{-1/2}
       \Sigma^{2/3},
\end{equation}
  where $\alpha_{\rm SS,-1}=\alpha_{\rm SS}/0.1$.
  From a pragmatic standpoint, even though the use of a thin
disk at small radii would be inconsistent because one would
  find $h/r >>1$, the use of a slim disk at small radii versus
a thin disk has minimal effect
  on the light curve. This comes about because the controlling time
scale in the problem comes from the largest 
disk radius.
 In fact, one could probably model the entire
  disk as a one-zone model, taken at the outer disk edge
  as was done by Metzger et al. (2008a)
  for NS-NS merger disks, and still get a reliable result.

\vskip -0.065 cm
\begin{figure}[h!]
\begin{center}
\includegraphics[height=130mm]{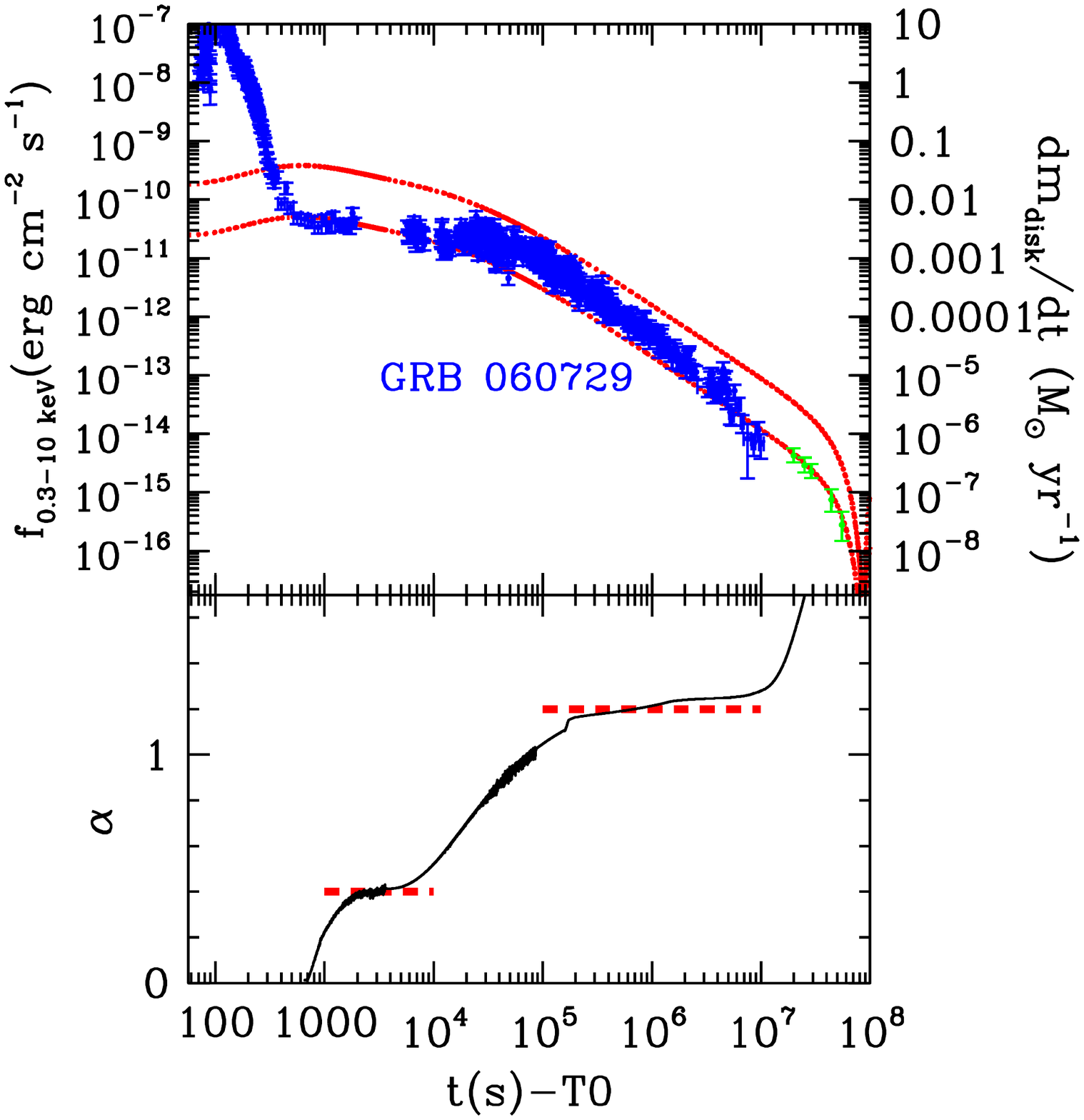}
\end{center}
\vskip -3.5 cm
\caption{
Model light curve
 showing the accretion-derived
   X-ray flux  from the fall-back 
   disk in a lGRB,
  taking
 $M_{\rm BH} =  5\msun$
  and $M_{\rm disk} = 10^{-4}\msun$.
  Also shown is the {\it Swift} XRT light curve (Evans et al. 
        2007, 
        2009)
  of GRB 060729 (in blue). 
 Given the value $4\pi d_L^2=1.15\times 10^{57}$ cm$^2$
 for  GRB 060729 ($z=0.54$),
   the efficiencies assumed in the two scaled 
 model light curves (in red) to convert
  from rate of accretion onto the
   central engine to
  XRT fluxes
     are $f\epsilon_{\rm net}^{-1}
          =8.8\times 10^{-3}$ (upper)
    and  $ 1.4\times 10^{-2}$ (lower).
     The accretion rate values given on
the right hand side axis correspond
to an overall efficiency $f\epsilon_{\rm net}^{-1}
          =4.9\times 10^{-3}$.
 The five flux
   values at late times $10^7$ s $<t<10^8$ s
   (in green)
   come from {\it Chandra} observations
  (Grupe et al. 
   2010). The last data point lies at $t-T0=642$ d.
  The initial profile is determined
  from Heger progenitor model 12SE
  as   explained in the text.
  The flat decay portion
  in the model light curve corresponds to the 
   time during which the 
  initial $\Sigma(r)$ 
  is being
redistributed into an accretion disk
    with a self-consistent radial profile.
   The lower panel shows the evolution of the locally 
defined rate of decay $\alpha=-d\log L_X/d\log t$.
   The two quasi-constant portions, $\alpha\simeq0.4$ and $1.2$,
  are indicated by the heavy dashed red lines. The long term
  decay rate $\sim1.2$ is well-characterized by the analytical
  solution for electron scattering, gas pressure 
   disks, $\alpha= 19/16$ (Cannizzo, Lee, \& Goodman 1990).
 }
\label{fig:2}
\end{figure}

The composition of the accretion
  disk is substantially different from 
   solar. The $T(\Sigma)$ relation exhibits 
   a weak inverse scaling with
  mean molecular weight $\mu$ for gas pressure and electron
scattering opacity, $T\propto \mu^{-1/3}$ (Cannizzo  \& Reiff 1992).
   The expected mixture of alpha elements
   yields $\mu\simeq 16$, rather than the value $\mu\simeq 0.65$
  more generally relevant for solar composition material.
 Thus $T\propto \mu^{-1/3}$ brings about a decrease in $T(\Sigma)$
  by a factor $\simeq2.9$ at given surface density,
   and therefore $h/r$ by a factor $\simeq1.7$,
  compared to solar composition material.
 
  For completeness, we note that significant outflow may
accompany accretion flows that are substantially super-Eddington.
  For computational expedience we only consider the slim
disk in the super-Eddington limit, since the addition of outflows
would introduce additional free parameters. However,
  given the strong potential for outflow, we
  stress that our accreted masses represent lower 
limits as to the amount of mass that may actually be accreted
  during  segments II and III.
  Similarly, the inferred accreted masses shown in Figure 1
 may significantly underestimate the true remnant masses
left over after the collapsar event.

\vskip -0.065 cm
\begin{figure}[h!]
\begin{center}
\includegraphics[height=130mm]{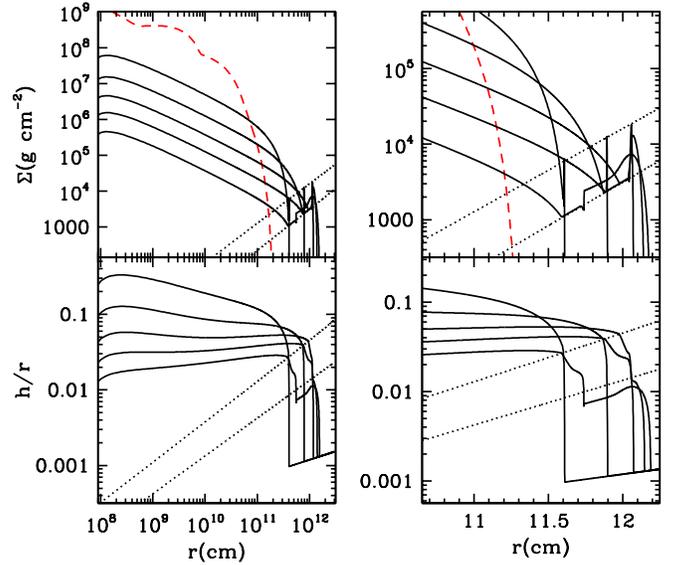}
\end{center}
\vskip -3.875 cm
\caption{
 The disk evolution accompanying   Figure 1.
    The upper panels show $\Sigma(r)$
    (solid lines,
    top to bottom)
  at
   $t=5$, 25, 100, 300, and 600 d.
  The dashed line
  indicates the initial $\Sigma(r)$ distribution,
  and the parallel dotted lines
  indicate the critical values associated with the
   dwarf nova limit cycle which occurs at the
   transition between  neutral and ionized gas.
  The lower panels show the evolution of $h/r$
  corresponding to the same times
  as in the upper panels.
}
\label{fig:3}
\end{figure}

     CG09   present
  analytical models for the afterglow light curves
 powered by fall-back disks.
  The main strength of model is the universal
 decay law $d\log L_{\rm acc}/d\log t \simeq -1.3$
  characterizing fall-back disks without the external,
confining tidal torque of a companion star
    (Cannizzo, Lee, \& Goodman 1990).
   This law seems consistent with the late time decay of GRBs
  seen in X-rays.
     The early time decay, in particular the steep
  decay for  $t\la10^3$ s, following the prompt emission,
   is difficult to explain 
   within the fall-back disk scenario and
     may be giving us information
     about the radial density profile within the progenitor (Kumar et al.
   2008b).
   In this work we present results of time dependent numerical 
   calculations using a general accretion disk code 
     (Cannizzo et al. 2010),
   which uses the input physics detailed in    CG09.
    The boundary conditions are as follows:
     (i)  at the inner disk  edge, taken to be
  $10^7$ cm for sGRBs and $10^8$ cm for lGRBs,
     matter is instantly
   removed as it arrives,  while
     (ii) the outer grid point is placed at such a large
   radius, typically $10^{13}$ cm, that during the course of the run the
   outer edge of the spreading fall-back disk never reaches it.
   The large dynamic range in disk radii $r_{\rm outer}/r_{\rm inner}$
  necessitates $\sim1500-3000$ grid points.
   In addition, no fresh material is added during a run. Thus the
evolution is set entirely by gradients within the disk, unlike the standard
  Shakura-Sunyaev disk fed at a constant rate in the outer edge  
    which approaches a steady-state ${\dot M}(r)=$ constant with time.

\section{LGRBs}


\vskip -0.065 cm
\begin{figure}[h!]
\begin{center}
\includegraphics[height=130mm]{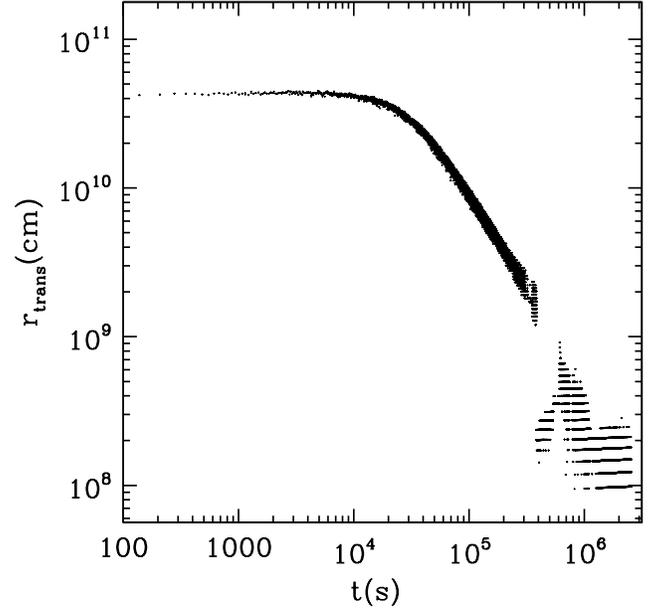}
\end{center}
\vskip -3.5 cm
\caption{
   The evolution of the boundary between the thick and thin
disk states, described in the text, for the 
   run shown in Figures 2 and 3.
   }
\label{fig:4}
\end{figure}

LGRBs are thought to be associated with 
    the explosions of 
   massive stars 
   (MacFadyen \& Woosley 1999, 
  Woosley \& Heger 2006).
  The initial failure of the SN
 is the reason
 for the GRB in the collapsar model.
  The hosts for lGRBs
   tend to be subluminous, irregular
  galaxies rich in star formation (Fruchter et al. 2006).
  If the long-term X-ray light curves of GRBs are indicative
 of feeding from a fall-back accretion disk, then from Figure 1
we see that no more than $\sim10^{-4}-10^{-3}\msun$
  of material in the fall-back disk is needed,
    for nominal assumptions about the efficiencies.
  In other words, for a  $\sim10\msun$ progenitor,
   a mass fraction only $\sim10^{-5}-10^{-4}$ 
   (excluding the $\sim10\msun$ which ends up within 
  $t \la 10-10^2$ s in the BH)
    is required to survive and persist in the vicinity
  of the progenitor to power the long-term light curve.
  If a significant fraction of the mass that tries
to accrete onto the BH is ejected,  the remnant mass
  around the BH could be substantially larger.
   We stress that in this discussion we only include mass
accretion beginning in segment II, since segment I is too
steep to be accounted for easily in the accretion disk
   scenario.

  Figure 2 shows a light curve for $M_{\rm BH} = 5\msun$
  and $M_{\rm disk} = 10^{-4}\msun$,
  overlaid with XRT data from GRB 060729.
   The initial $\Sigma(r)$ profile is taken
  from Heger progenitor\footnote{models
  available at {\tt http://2sn.org/GRB2}}
     12SE
   (Woosley \& Heger 2006)
in the following way:
 The progenitor is collapsed in cylindrical coordinates
along the $z$-axis to obtain $\Sigma_{\rm initial}(r)$,
   and then scaled down in mass 
  with a radial factor $\propto(r/R_*)$ so that 
$\int 2\pi r dr \Sigma_{\rm initial}(r)$ equals $10^{-4}\msun$.
 A radial scaling is also applied so that the integrated
specific angular momentum in the final state, 
taken to represent the disk and therefore to be Keplerian,
equals that in the progenitor. This 
  effectively introduces  a radial
scaling $\sqrt{j_{\rm tot}/j_{\rm crit}}$
  to the $\Sigma_{\rm initial}(r)$ distribution.
Due to the very long plateau of GRB 060729,
we adopted a progenitor model with a large radius.
 (The value $\log t^*_{\rm II}({\rm s}) \simeq 4.7$ 
  for GRB 060729 is at the upper end of the 
 distribution of $\log t^*_{\rm II}$ values given
in Dainotti et al. 2010.)
   The lower panel of Figure 2 indicates the
 local value of the temporal decay index
  $\alpha$ associated with the model.
 There are two power law segments
  evident in the model decay light curve,
   a brief one $\alpha\simeq0.4$ 
   between $10^3$ and $10^4$ s,
  and a longer one $\alpha\simeq1.2$
     between $10^5$ and $10^7$ s.
 The latter value $\alpha\simeq1.2$ is close to that expected
 from analytical models, $19/16$ (Cannizzo, Lee, Goodman 1990).
 Figure 3 gives the evolution of surface density $\Sigma(r,t)$
  and local aspect ratio $h/r$, and Figure 4 indicates the
  transition point between the slim disk and thin disk states,
i.e., $h/r$ dropping below $(3/4)^{1/2}$.

  The steep decay associated with segment I
 is not accounted for in our model, and may well
 be due to a separate physical process,
 such as the accretion of the progenitor core 
 (Kumar et al. 2008b; 
 Lindner et al. 2010,
  see their Fig. 2).
  The time scale $t_{\rm II}$
 for the plateau 
associated with segment II
   comes out naturally in
 the models, given the $\sim10^{11}$ cm radius for the progenitor
   $-$ 
   assuming that the initial fall-back mass distribution 
   roughly traces the envelope of the progenitor
(after taking into account the radial
  scaling factor $\sqrt{j_{\rm tot}/j_{\rm crit}}$).
    The observed X-ray decay for GRB 060729,
  the GRB which has the been 
   observed for the longest time
  in X-rays (Grupe et al. 2007, 
                         2010),
     does not match precisely
   that of the fall-back disk.  
   There may be various systematic
  effects  that could account for
 the difference.
  The most obvious effect,  
    a variable cosmological K-correction 
  concomitant with
   the fading in X-ray flux over 
  a dynamic range of $\sim6$ decades,
 cannot play a strong role,
   as
   the spectral index does not vary
  significantly
   over the long term
  from its mean value $\beta\simeq 2$.
   There may be some
  time-variable physics associated
 with the X-ray emission from within the jet
  that could affect either $\epsilon_{\rm net}$  
  or $f$ versus time. 
    Our calculations only represent the accretion
  disk which powers the jet
  and we take  
      ${\dot \epsilon}_{\rm net} = {\dot f} = 0 $ for 
   simplicity.
   Modeling the jet emission is beyond the scope of this work.

At late times $10^7$ s $<t<10^8$ s
 there is a deviation from the canonical power law decay
 because  $\Sigma(r_{\rm outer})$
    drops to the point at which the dwarf nova limit
   cycle becomes active (indicated by the
  parallel dashed lines in Figure 3),
    and a cooling front is launched from the
   outer edge that instigates a transition from ionized to
  neutral gas. This quenches the source of accretion onto
  the central engine.
     The late time {\it Chandra} observations of 
   GRB 060729  (Grupe et al. 2010)
    appear to  coincide with the timing of this
  drop-off, which may therefore represent a cooling transition
  front in the outer disk which diminishes the rate of 
   accretion onto the central engine, rather than a jet break.

\section{SGRBs}


SGRBs are thought to be caused
   by the merger of two neutron stars
 (Eichler et al. 1989;
  Paczy\'nski    1991;
  Narayan, Piran, \& Kumar 2001;
  Rosswog \& Ramirez-Ruiz  2002).
  The hosts for sGRBs 
    are indicative of an older
  population with less active star formation
   (Fox et al. 2005;
    Gehrels et al. 2005;
    Villasenor et al. 2005;
    Bloom et al. 2006),
  and also
 tend to be $\sim5$ times
  more spatially extended than those 
 of lGRBs,
 matching the $\sim5$ times wider
  projected spatial 
   distribution of sGRBs vs. lGRBs
   within their hosts
  (Fong, Berger, \& Fox 2010).
 Due to their faintness relative to lGRBs, sGRBs
  are much less well-studied in X-rays.
  Indeed, there is only one certifiable example of
  a sGRB  which has a long and well-sampled X-ray light curve,
  to the extent that a statement can be made as to
 the presence of a plateau -  GRB 051221A.

\vskip -0.065 cm
\begin{figure}[h!]
\begin{center}
\includegraphics[height=130mm]{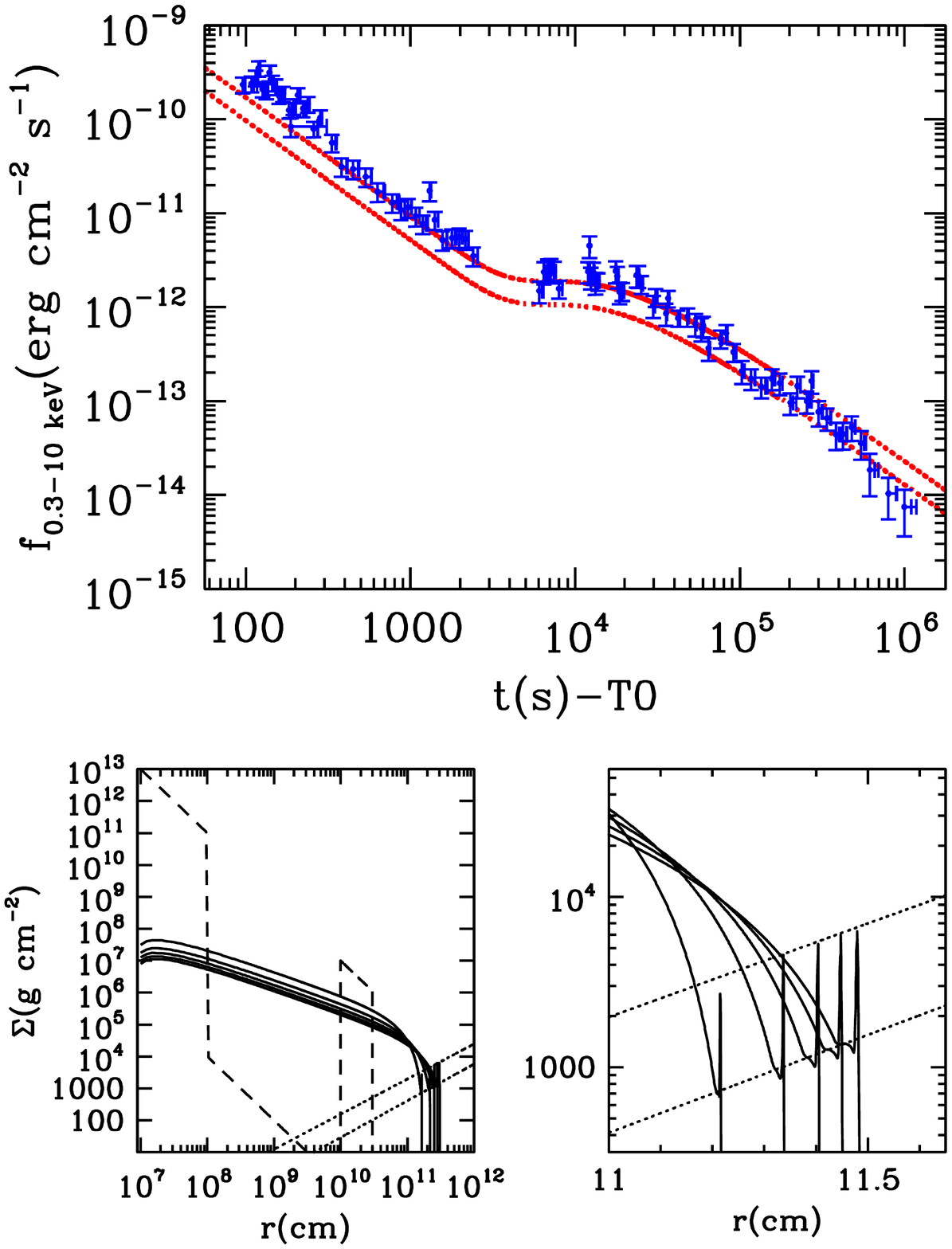}
\end{center}
\vskip -2.05 cm
\caption{
  Model light curve showing the accretion-derived
    jet power from the fall-back disk in a 
 sGRB, taking $M_{\rm BH} = 3\msun$
  and $M_{\rm disk} = 10^{-5}\msun$.
   Also shown is the  {\it Swift} XRT light curve 
    (Evans et al. 2007, 
                  2009)
   of GRB 051221A (in blue).  
 Given the value $4\pi d_L^2=1.18\times 10^{57}$ cm$^2$
 for  GRB 051221A ($z=0.5465$),
   the efficiencies assumed in the two scaled 
 model light curves (in red) to convert
  from rate of accretion onto the central engine to
  XRT fluxes    are $f\epsilon_{\rm net}^{-1}
     = 2.4\times 10^{-2}$ (upper)
  and $4.3\times 10^{-2}$ (lower).
    One sees fairly rapid decay initially
   due to the presence of a significant amount of accreting
   material at small radii. The brief shoulder in the
   light curve is due to the delayed accretion of the spiral
arm of ejected NS matter at large radii, motivated by
detailed SPH calculations (Rosswog 2007).
   The bottom panels show the surface density
  evolution, at equally spaced intervals $\delta t = 2$ d.
   The dashed line indicates the initial $\Sigma(r)$ distribution,
  and the parallel dotted lines
  indicate the critical values associated with the  dwarf nova limit cycle.
       }
\label{fig:5}
\end{figure}

The importance of the outward spreading of the disk
   formed from the NS-NS merger was clearly demonstrated
  by Metzger et al. (2008a,
    see their Fig. 1).
  The initial radial density profile,
  and hence $\Sigma(r)$ profile after after angular momentum
  conservation has vertically compacted the merger remnants into 
their accretion plane, is of course much more radially
  condensed
  than that expected for lGRBs.
      Theoretical guidance on a potential $\Sigma(r)$
  profile to begin the calculations,
  particularly at large radii, is scarce, 
   yet there are indications of small amounts of
  matter $\sim10^{-6}-10^{-4}\msun$ ejected at early times
   that may have high angular momentum, and 
  therefore circularize at large radii 
   (Rosswog 2007).  One potentially important effect
we cannot model is the disruption of a portion of the disk
  by strong nucleosynthesis in the NS supplied material
  as it expands rapidly to subnuclear densities
  (Metzger et al. 2008a, 2008b, 2010).
  Lee, Ramirez-Ruiz, \& L\'opez-C\'amara
  (2009) find that strong winds can be 
launched from the surface of post-merger
NS-NS disks, powered by the recombination
 of free nucleons into $\alpha-$particles.

Figure 5 shows the evolution of a fall-back disk
  of potential relevance
  for the aftermath of a NS-NS merger.
  We take  $M_{\rm BH} = 3\msun$
  and $M_{\rm disk} = 10^{-5}\msun$.
  Unlike the 
  much longer evolution shown in Figure 2 for lGRBs,
  in this case  the  outer edge of the
  disk is still freely expanding to larger radii
 by the end of the run
  (indicated by the narrow spike in $\Sigma(r)$ between
 $10^{11}$ cm and  $3\times 10^{11}$ cm
  shown in the second lower panel).
   The brief plateau at $\sim10^4$ s
   results from the ad hoc introduction of 
  the small amount $\sim 10^{-9}\msun$
  of high angular momentum material at the 
  large circularization radius $10^{10}$ cm.
   The high efficiencies
   associated with
      BH accretion  show the potential for
  a small amount of material $\ll 1\msun$
  to have  a dramatic effect on the long-term light 
  curves
   as regards brief plateaus or inflection points.

\smallskip
\smallskip

\section{Potential Application to a Dainotti-like relation}
 
\smallskip
\smallskip

In Section 2 we show that the Dainotti relation
  $L^*_{\rm II} \simpropto  {t^*}_{\rm II}^{-1}$
   as originally envisioned may be due to the  observational
bias against detecting and characterizing faint plateaus.
    Nevertheless, the accreted mass
  estimates $\delta M$
   inferred
   using the $L^*_{\rm II}$ and $t^*_{\rm II}$ values 
   from Dainotti et al. (2010)
are interesting in the context of this
   work, and one might legitimately ask what theoretical
   prediction 
   for  the $L^*_{\rm II}(t^*_{\rm II})$ relation 
the fall-back accretion hypothesis would make, given
     a hypothetical physical constraint in which $\delta M$ were
held constant and the initial radius of the fall-back disk were allowed to vary.
Figure 6 shows the results of seven runs in which we fix $\delta M= 10^{-4}\msun$
   and vary the initial outer radius of the fall back disk.
   One can see a clear inverse relation between the duration of the plateau
  $t^*_{\rm II}$ and the luminosity at the end of the plateau $L^*_{\rm II}$.
    The low$-z$ behavior of $\delta M$ seen in Figure 1, however,
   indicates that a spread in $\delta M$ may be more realistic, in which case
  one would not expect to be able to use the theoretical prediction
   of the $L^*_{\rm II}(t^*_{\rm II})$ relation 
as a useful discriminant for the theory.

\vskip -0.065 cm
\begin{figure}[h!]
\begin{center}
\includegraphics[height=130mm]{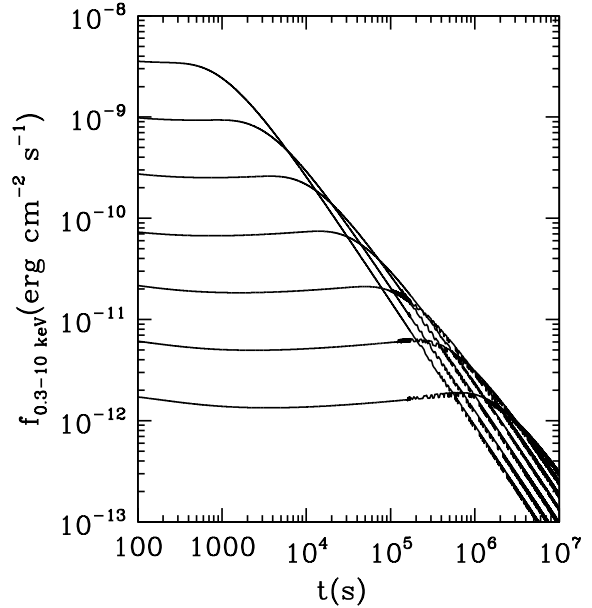}
\end{center}
\vskip -3.95 cm
\caption{
  Model light curves for lGRB parameters, keeping
  the initial fall-back disk mass constant at $10^{-4}\msun$
   but varying the initial radius and overall normalization.
  For the central light curve $r_0=7\times 10^{10}$ cm.
  For each successive light curve $r_0$ is increased a
   factor of two going to the right, and decreased a
factor of two going to the left.
   To keep $\delta M$ constant, the overall normalization
  on $\Sigma(0)$ is varied by an additional factor of 2.5
   for each successive run. 
       }
\label{fig:6}
\end{figure}

\section{Discussion and Conclusion}

\smallskip

We have presented time dependent calculations of the fall-back
   disk scenario to account for the long-term X-ray light curves
  for GRBs. For lGRBs, an initial radial scale of a 
 stellar radius $\sim10^{11}$ cm
gives a natural viscous evolution time 
   of $\sim10^4$ s to redistribute the
matter into a quasi-steady disk, roughly consistent with the 
      observed plateau duration.  GRB 060729
  which we model in the context of a lGRB
  had one of the longest plateaus ever observed,
 thus more generally we anticipate
    progenitor radii, or more specifically,
circularization radii, given by $\sim \sqrt{j_{\rm tot}/j_{\rm crit}}R_*$,
to  lie in the range $\sim 10^{10} - 10^{11}$ cm.
  The rate of decay for GRB 060729
  is close but  does not match the models
  precisely, which may hint at 
  time-variable emission processes that would affect
  our adopted efficiencies $\epsilon_{\rm net}$ and $f$.
It is interesting to note that our effective
  values of the efficiencies, accretion plus beaming,
  required to match the observed X-ray flux levels
  are comparable between lGRBs and sGRBs:
  $f \epsilon_{\rm net}^{-1} \simeq 10^{-2}$ for 
   GRB 060729  and 
  $f \epsilon_{\rm net}^{-1} \simeq 3\times 10^{-2}$ for 
   GRB 051221A.
  If the accretion efficiency $\epsilon_{\rm net}$ is about the same for
  lGRBs and sGRBs, this may indicate that sGRBs are
  less beamed by a factor of  $\sim3$ compared to lGRBs, 
  roughly in line with previous results 
  (Watson et al. 2006, 
     Grupe et al. 2006).
     This similarity 
   between lGRBs and sGRBs 
  afterglows
is not entirely unexpected, given
 the general similarities in their afterglow
  properties (Gehrels et al.                 2008;
              Nysewander, Fruchter, \& Pe'er 2009).

The concept of powering the long term X-ray
light curve of GRBs by accretion onto the BH represents
 a departure from the standard model in which the
  fading corresponds to the deceleration of
 a baryonic jet.  In the accretion model,
  the jet itself would be  very light,
   perhaps composed almost exclusively 
  of Poynting flux, and the long term
  fading would not be due to variations
  in the Lorentz factor  and beaming factor.
 Recent high fidelity GRMHD calculations 
  of BH accretion support the idea of
a high Lorentz factor jet with minimal   baryon loading
  (McKinney \& Narayan 2007ab).
  The accretion scenario involving fall-back of the
progenitor core has been examined in detail by Kumar et al. (2008a, 2008b)
and Lindner et al. (2010).  Kumar et al. (2008b) argue that
   constraints may be placed on the density profiles and radii
of the progenitor core and envelope, as well as their
rotation rates relative to break-up.
  Lindner et al. (2010) examined the progenitor core fall-back
scenario in much greater detail with 
  the  adaptive
mesh refinement FLASH code,
  where their calculations are done in cylindrical coordinates.
  Starting with a Heger progenitor, they follow the
  fall-back evolution of the progenitor. 
Their Fig. 2 shows the potential for the model to
obtain a steeply decaying light curve as is seen in X-rays
  for segment I, where $\alpha\simeq3$ in the
Zhang et al. (2006) schematic.
  In their Discussion, Lindner et al.  mention
several caveats, such as the lack of nuclear physics
  in the inner disk, the neglect of the MRI, and the
lack of modeling the axial relativistic jet.
   A more basic concern is simply the range of rate of
accretion in comparison to that inferred from 
observations. Fig. 2 of Lindner et al. shows that
  if progenitor core fall-back is the correct explanation
for segment I in the X-ray light curve, the rate of
accretion onto the BH during this phase varies between
about 0.1 and $10^{-5}\msunsec$, whereas from Fig. 2 of this
work we see that, for nominal assumptions about the accretion
efficiency, the rate inferred from observations 
  on segment I for GRB 060729
   varies
  between about 10 and $3\times 10^{-3}\msunyr$, 
lower by a factor $\sim10^5-10^6$ than in Lindner et al. (2010).
  One does in fact expect for the theoretical accretion
    rate in this context to be an overestimate,
   given the potential for outflow,
but the discrepancy here seems extreme.
   There are several
possibilities for this discrepancy. The vertical scale shown in 
  Fig. 2 of this work is already super-Eddington by between 
about one and eight orders of magnitude, and therefore 
the range in the Lindner et al. calculations would be 
correspondingly greater.  It may be that even highly advective
disks cannot accommodate
that much accretion onto the BH, and that a significant fraction
 of the material is blown away before it can accrete.  This would
have to occur, however, in such a way that it did not 
   interfere with
the propagation of the jet. Also, for the shape
  of the decay light curve calculated by Lindner et al. to correspond
to segment I, the ratio of accreted to ejected gas would have to
 remain constant. If it varied significantly, the accretion-derived
   luminosity would have a different decay power law.
 Another possibility is that
the Lindner et al. calculations greatly over-estimate the fall-back,
  in which case a much greater fraction of the progenitor would be
ejected on a time scale $\la10^3$ s.   Lindner et al. note that they
do not find evidence for the thin disk hypothesized by CG09 as
  characterizing the fate of the progenitor envelope fall-back.
    Given the apparent mismatch in accretion rates between their theory
 and the observations, and the findings of this work, their 
criticism could have at least two mitigating factors:
(1) In Figure 4 we see that for the period covered by the Lindner 
et al. calculations, namely $t<10^3$ s, a significant fraction of the
disk lies within $r_{\rm trans}$, i.e., the more spatially 
  extended slim disk rather than the 
thin disk. 
(2) Given the apparent mismatch between theory and observation for the
   rates of accretion, the actual densities within the volume 
formerly occupied by the progenitor may be much less than calculated
  by Lindner et al. (2010),
  which would interfere far less
  with an accretion disk formed from fall-back debris.

   At late times {\it Chandra} observations indicate a steepening  in the
rate of decay (Grupe et al. 2010), which appears 
   to be consistent with the onset of a cooling front
  in the disk. This would be an alternative to the standard jet-break interpretation
    discussed by Grupe et al. (2010).
   For sGRBs, the picture is less clear, given that we only have a single well-studied
    example. We have shown that the presence of even a very small amount of
  high angular momentum gas $\sim10^{-9}\msun$ can give  a slight inflection to the 
X-ray decay, as was observed in GRB 051221A.
  As for the overall X-ray light curve,
  segments II and III,
   if only $\sim 10^{-5} - 10^{-4}\msun$ of gas
   survives either the  hypernova (lGRBs) or NS-NS merger (sGRBs), 
    then the accretion resulting from the ensuing fall-back disk
  should power a long-term jet.

  Lastly, we have shown that the Dainotti relation 
   $L^*_{\rm II} \simpropto  {t^*}_{\rm II}^{-1}$
  may be due to an observational bias against detecting and 
  characterizing faint plateaus:  the relation
  is governed by GRBs at $z\ga1.5$ for which we only detect
 the upper envelope of a broad distribution.
    Nevertheless,  the existence 
  of an apparent upper limit to the 
  total X-ray energies inferred from the X-ray fluences, 
  and therefore the accreted masses if one assumes accretion onto
  the central engine as the long term powerhouse for the X-ray flux,
   is extremely interesting.  For nominal values of
   the accretion efficiency $\epsilon_{\rm net}$ and
  the beaming factor $f$,
    we find an upper limit $\simeq 10^{-4} - 10^{-3} \msun$
     for the accreted  mass
 during segments II and III.
 (The lower end of the distribution
  of accreted masses,
   which is partially revealed 
 for GRBs at  $z\la1.5$,
 may extend down to $\simeq  
10^{-8} - 10^{-7} \msun$.)
  This  means that for a progenitor
  mass $\sim10\msun$, only a maximum mass fraction $\sim10^{-5} - 10^{-4}$
   of the progenitor survives in the vicinity of the progenitor
  to be accreted as a fall-back disk  (excluding the $\sim10\msun$
   that ends up in the BH during the prompt emission
    and subsequent segment I
  consisting of the steep-decay).
  This has important ramifications for the energetics associated 
  with the  hypernova explosion and subsequent removal of most 
  of the progenitor envelope.

\smallskip

We acknowledge useful conversations 
          with Maria Dainotti, 
                Dirk Grupe,
             Stephan Rosswog,
          and Brad Schaefer.
 This work made use of data supplied by the UK {\it Swift} Science Data Centre
   at the University of Leicester.

\end{document}